\documentclass[12pt]{amsart}
\usepackage{latexsym,fancyhdr,amssymb,color,amsmath,amsthm,graphicx,listings,comment}

\usepackage[section]{placeins}
\let\paragraph\subsection
\title{Positive curvature and bosons}
\author{Oliver Knill}
\date{06/28/2020}
\address{Department of Mathematics \\ Harvard University \\ Cambridge, MA, 02138 }

\begin{document}

\begin{abstract}
Compact positive curvature Riemannian manifolds M with symmetry group G allow Conner-Kobayashi 
reductions M to N, where N is the fixed point set of the symmetry G. The set N is a union
of smaller-dimensional totally geodesic positive curvature manifolds each with even co-dimension.
By Berger, N is not empty. By Lefschetz, M and N
have the same Euler characteristic. By Frankel, the sum of dimension of any two components 
in N is smaller than the dimension of M. Reverting the process N to M allows to build 
up positive curvature manifolds from smaller ones using {\bf division algebras} and the geodesic flow.
From dimension 6 to 24, only four exceptional manifolds have appeared so far, some of them
being flag manifolds and related to the special unitary group in three dimensions. We can now draw a periodic
system of elements of the known {\bf even-dimensional positive curvature manifolds} and
observe that the list of even-dimensional known positive curvature manifolds has 
an affinity with the list of known {\bf force carriers in physics}. 
Positive mass of the boson matches up with the existence of of two
linearly independent harmonic k-forms on the manifold.
This motivates to compute more quantities of the exceptional positive curvature 
manifolds like the lowest non-zero eigenvalues of the {\bf Hodge Laplacian} L=d d*+d* d
or properties of the pairs (u,v) of harmonic 2,4 or 8 forms in the positive mass case. 
\end{abstract}

\maketitle

\paragraph{}
Similarly as {\bf representation theory} lets {\bf Lie groups} act on linear spaces, one
can see some Lie groups realized as symmetry groups acting as isometries on 
{\bf compact even-dimensional positive curvature manifolds}.
So far, in even dimensions, only spheres $\mathbb{S}^{2d}$, projective spaces of division algebras 
$\mathbb{RP}^{2d},\mathbb{CP}^d,\mathbb{OP}^2$ as well as the four special manifolds
$W^6,E^6,W^{12},W^{24}$ are known to admit positive curvature. All of them also admit a continuum
isometry group. 

\paragraph{}
The exceptional manifolds are the {\bf Wallach manifolds} $W^6,W^{12},W^{24}$ of 
complete flags in three-dimensional \footnote{These are 2*3=6,4*3=12,8*3=24 dimensional real vector spaces.} 
vector spaces $\mathbb{C}^3,\mathbb{H}^3, \mathbb{O}^3$ over the
{\bf division algebras}
$\mathbb{C},\mathbb{H}, \mathbb{O}$ and the {\bf Eschenburg manifold} $E^6$ which is a twisted 
version of $W^6$ with same cohomology but different cohomology ring \cite{Ziller2}. The manifold $E^6$ is
a {\bf bi-quotient}, where $(z_1,z_2) \in \mathbb{T}^2$ acts on $SU(3)$ as $g \to z_1 g z_2^{-1}$.
These exceptional $2d$-manifolds differ from the {\bf projective spaces} of division algebras in 
that some of their {\bf inner Betti numbers} $b_k(M)$ are larger than $1$. 

\paragraph{}
An affinity to {\bf particle physics} appears:
the {\bf force carrier bosons} $W^{+},W^{-},Z^0,H$ ({\bf vector gauge bosons} and 
the {\bf scalar Higgs boson} $H$) 
which all have {\bf positive mass} can be lined up with the exceptional positive curvature manifolds 
$W^6,E^6,E^{12},W^{12}$ with ``heavier" {\bf cohomology},
the projective spaces match with the rest: complex projective spaces with {\bf photons},
quaternionic projective spaces with {\bf gluons} and the {\bf Moufang-Cayley plane} 
$\mathbb{OP}^2$ with the {\bf graviton}, a particle which is not expected to be discoverable 
with current technology \cite{RothmanBoughn}.

\paragraph{}
It appears that one can look at compact even-dimensional positive curvature manifolds $M$ 
constructively, starting with the $0$-manifold $1=\mathbb{RP}^0$ and build them with 
successive extensions $N=N_1 \cup \cdots \cup N_k \to M$ using a few principles: 
{\bf 1. Conner-Kobayashi:} extensions must come from a Lie group symmetry.
\cite{Conner1957,Kobayashi1958},
{\bf 2. Frobenius-Hurwitz:} extensions come from unit spheres $Z_2,U(1),SU(2),S^7$ of 
                            $\mathbb{R},\mathbb{C},\mathbb{H},\mathbb{O}$
\cite{Frobenius1878,Hurwitz1922,Lam1991} (Frobenius (1877) assumes an associativ division algebra and gives 
$\mathbb{R},\mathbb{C},\mathbb{H}$, Hurwitz (1923 posthumous) assumes normed division algebra and gives all 4),
{\bf 3. Moufang-Cayley:} only one single $S^7$ extension is possible, the Moufang-Cayley 
plane is the last. \cite{Moufang33,Salzmann1995},
{\bf 4. Frankel-Synge:} the fundamental group is $\mathbb{Z}_2$ or $0$ and
there is a dimension inequality from variational principles.  \cite{Frankel1961,Synge},
{\bf 5. Euler-Lefschetz:} Euler characteristic is invariant under extensions. 
\cite{Kobayashi1972,Petersen3}. A guiding principle is therefore to place the manifolds
in the {\bf dimension-Euler characteristic} plane which then has the structure of a 
{\bf periodic system of elements}.

\paragraph{}
Writing $M_{\chi(M)}$, where $\chi(M)$ is the Euler characteristic,
the list is $\mathbb{RP}^{2d}_1,\mathbb{SP}^{2d}_{d+1},\mathbb{CP}^d_{d+1},\mathbb{HP}^d_{d+1}$  for all $d$
and $\mathbb{OP}^d_{d+1}$ for $d=1,2$, the Wallach manifolds \cite{Wallach1972} 
$\mathbb{W}^6_6,\mathbb{W}^{12}_6,\mathbb{W}^{24}_6$
and the Eschenburg manifold \cite{Eschenburg1982} $\mathbb{E}^6_6$. See \cite{Ziller2} for
an overview or \cite{Shankar1999} for symmetry groups, which are surprisingly large. 
Both the $W^{24}$ or the Moufang-Cayley plane $\mathbb{OP}^2$ have the symmetry of $F_4$
an exceptional simple Lie group \cite{Baez2002}.

\paragraph{}
The sequence of $\mathbb{CP}^d$ with $U(1)$ extensions are {\bf electro-magnetic}
in nature and correspond to {\bf photons}, the $\mathbb{HP^d}$ with $SU(2)$ extensions are 
{\bf electro-weak} and correspond to {\bf gluons}, the $\mathbb{O}^d$ are 
{\bf gravitational}. While $\mathbb{O}^1=\mathbb{S}^8$, the Moufang-Cayley plane $\mathbb{O}^2$
would correspond to the {\bf graviton} $G$, a spin 2 particle.
The $\mathbb{W}^6,\mathbb{E}^6,\mathbb{W}^{12}$ are constructed from $SU(3)$ and so have an affinity
with the {\bf vector bosons} $W^+,W^-,Z^0$. The $\mathbb{W}^{24}$ is linked to all type of symmetries
and also to the gravitational $\mathbb{O}^2$ and so naturally associated with the {\bf Higgs Boson}.
The link of division algebras with fundamental physics had been suggested 
already in 1934 by Jordan, von Neumann and Wigner \cite{JordanNeumannWigner,Baez2002}.

\begin{figure}[!htpb]
\scalebox{0.25}{\includegraphics{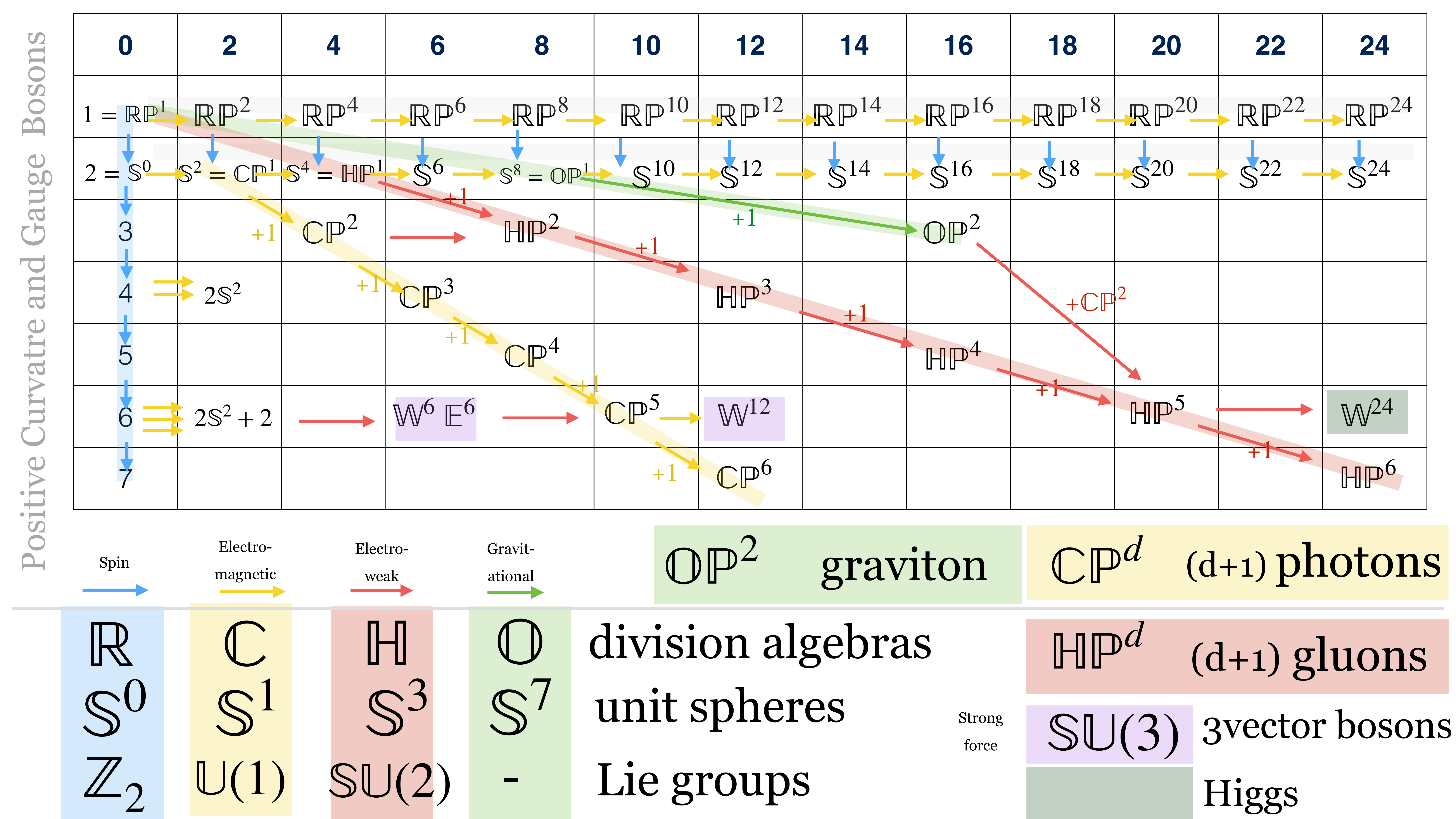}}
\label{bsosons}
\caption{
Periodic system of positive curvature manifolds. 
}
\end{figure}

\paragraph{}
We do not know whether the currently known {\bf ``periodic system of elements"} 
for positive curvature manifolds is complete.
The {\bf standard model of particle physics} also provides a system of fundamental forces and their bosons. 
Also there, we do not know whether it is complete. In this note, we suggest a remarkable affinity
between these two seemingly unrelated classification problems. This association forces to think 
about the nature of the fundamental forces in physics. This is not a new idea. Atiyah 
in a talk\cite{Atiyah2010} speculated about the possibility of associating octonions with gravity.
See also \cite{Gursey,DixonDivisionalgebras,Baez2012} and 
\cite{ConwaySmith,Baez2002,Furey2016,RowlandsRowlands}.

\paragraph{}
There are areas in mathematics, where one has complete classifications of objects like for 
{\bf finite simple groups}, or {\bf Lie groups}, for 
{\bf regular polyhedra} or for {\bf division algebras}. Similarly, in physics, one has 
a complete classification of say crystal structures related to wallpaper groups,
a classification of eigenvectors of the Hydrogen operator leading to 
the periodic system of elements in chemistry. 
There are also areas in mathematics, where a classification of objects
is missing, like compact positive curvature manifolds, four manifolds or a 
classification of knots. 

\paragraph{}
{\bf Symmetries} are important in any classification. They are related to {\bf conserved 
quantities} by Noether's theorem and to {\bf variational principles} like 
for example in the isoperimetric inequalities. 
Classification and cross field associations always come with risks. One can also get 
``lost in math" \cite{LostInMath}. Kepler's {\bf Harmonices Mundi} of 1618 linked polyhedra 
with planets in the solar system, William Thomson (Baron Kelvin) imagined in 1870 
{\bf a vortex theory} modeling atoms using knots in the aether. More examples are in \cite{Livio2013}.
In the representation theory of Lie groups, affinities with particle physics has worked out.
See \cite{Dosch2008} for the history. Some less like \cite{Golomb} seeing Baryons
and mesons in the Rubik cube, or \cite{ParticlesPrimes} for seeing Leptons and Hadrons
structures in equivalence classes of primes over division algebras $\mathbb{C},\mathbb{H}$. 

\paragraph{}
The case of {\bf real normed division algebras} $\mathbb{R},\mathbb{C},\mathbb{H},\mathbb{O}$
is even more remarkable than the finite list of {\bf Platonic solids}. 
According to the {\bf Hurwitz theorem}, the real complex numbers, quaternions and the 
octonions are the only real normed division algebras. The associated projective spaces are 
equivalence classes of lines in these algebras. 
The non-associativity of the octonions alters the projective construction.
In the associative cases, there are infinite 
sequences $\mathbb{RP}^{2d},\mathbb{CP}^d,\mathbb{HP}^d$ of even-dimensional 
projective spaces, coming from extensions $U(1),SU(2)$. 
The Moufang-Cayley plane $\mathbb{OP}^2$ however is the last octonionic one, the reason being
non-associativity \cite{lackmann2019}. One can see $Z_2$-extension as a passage going from projective spaces to spheres. 
But this is only needed for $\mathbb{RP}^{2d}$, as $\mathbb{CP}^d,\mathbb{HP}^d,\mathbb{OP}^d$ are
orientable and do not require a double cover. Universal covers can often linked to {\bf spin} 
like in the cover $SU(2) \to SO(3)$. 

\paragraph{}
All projective spaces over division algebras are compact manifolds all allowing
for positive curvature metrics with symmetry. Similarly as in the classification of 
regular polytopes, there are cases with infinite families like the 
$1$-dimensional polyhedra, regularity in the larger dimensional case as
in dimension 4 and higher only the higher-dimensional tetrahedra (simplices), 
cubes (hypercubes) and octahedra (cross polytopes) exist,
in dimension 3 and 4, a richer structure of two or three {\bf exceptional 
polyhedra} exist. In the case of polyhedra, the structure is determined 
by combinatorial constraints like the {\bf Gauss-Bonnet theorem}.

\paragraph{}
The known even-dimensional positive curvature manifolds are four infinite
sequences: {\bf spheres} and {\bf projective spaces}
$\mathbb{RP}^{2d},\mathbb{SP}^{2d},\mathbb{CP}^d,\mathbb{HP}^d$ 
then $\mathbb{O}^d$ for $d=1,2$ over the four division algebras. 
The ``rare elements" are the {\bf Wallach manifolds} 
$\mathbb{W}^6,\mathbb{W}^{12},\mathbb{W}^{24}$ and the 
{\bf Eschenburg manifold} $\mathbb{E}^6$. No other example has been found. It is also 
remarkable that all of these examples admit metric with symmetries. One can constrain
the situation more and ask at least one component $N$ to have co-dimension $2$. 
This has been completely classified by Grove-Searle \cite{GroveSearle}. 
The co-dimension-$4$ case with $SU(2)$ symmetry adds $\mathbb{HP}^d$. An 
example of a co-dimension $6$ case is $N=W^6$ for a $\mathbb{T}^1$ action on $M=W^{12}$.
\footnote{Personal communication Burkhard Wilking}.

\paragraph{}
Why should positive curvature manifolds and physics be linked at all?
Lie groups have to have a home somewhere. Representation theory lets them work on 
linear spaces. If we want to have them act on a compact space, then positive curvature 
is appealing as there is a {\bf reduction theory} similarly as in representation theory.
One can build up positive curvature manifolds $M$ from smaller dimensional ones $N$ which are
fixed point sets of a group $G$ action. It is nice that this reduction is always non-trivial
because of {\bf Berger's theorem} \cite{Berger1961} assuring that the 
fixed point set is non-empty in even-dimensions. This is not the case for negative curvature, where 
by the {\bf Cartan-Hadamard theorem}, there is no such theory. 
Still, both positive or negative curvature case could be relevant as the relativistic
constant curvature models {\bf de Sitter} and {\bf anti-de Sitter} illustrates. 

\paragraph{}
Many more questions remain beside the mathematical question to complete the 
list of positive curvature manifolds with or without symmetry.
In order to find out whether the affinity between positive curvature manifolds 
and force carriers in physics is just a structural accident, 
one should try to link the {\bf mass} $91.19 {\rm GeV}/c^2$ of the $Z^0$ boson 
with properties of $W^{12}$ or the mass $80.39 {\rm GeV}/c^2$ of $W^{\pm}$ bosons 
with the Wallach or Eschenburg manifold $W^6,E^6$ and the mass $125.35 {\rm GeV}/c^2$ 
of the Higgs boson with properties of $W^{24}$. 
Beside dimension and Euler characteristic, {\bf volume},
{\bf diameter} and Betti vectors, properties of harmonic forms or {\bf spectra}, 
especially ground state energies offer themselves.

\paragraph{}
We have $b(W^6)=b(E^6)=(1,0,2,0,2,0,1)$ (with different cohomology ring) 
\cite{Ziller2} or $b(W^{12})=(1,0,0,0,2,0,0,0,2,0,0,0,1)$,  \\
as well as $b(W^{24})=(1,0,0,0,0,0,0,0,2,0,0,0,0,0,0,0,2,0,0,0,0,0,0,1)$. 
\footnote{Personal communication Wolfgang Ziller}.  
The $b_0=b_{2d}=1$ are not of interest in connected and orientable cases.
The interior Betti numbers are associated with harmonic forms by {\bf Hodge theory}.
They can be represented by varieties or even manifolds 
with properties like volume or intersection numbers.

\paragraph{}
The picture appears that the geodesic flow on $W^6$ and $E^6$ are very 
similar, just introducing different oriented return maps, obtained by starting the 
geodesic flow on one sphere $\mathbb{S}^2$ of the symmetry and ending on the second. 
It is the only case where we see two positive dimensional components $N_1,N_2$ in $N$.
By Frankel \cite{Frankel1961}, 
$2+2={\rm dim}(N_1) + {\rm dim}(N_2) \leq {\rm dim}(M)-2=4$. This Frankel inequality
is hard to meet in higher dimensions as it forces extensions $N \to M$
which go beyond the known list of division algebras. 
All even-dimensional positive curvature manifolds appear to admit metrics with
{\bf integrable geodesic flow}. This can be understood by looking at the return properties
of the flow outside $N$ and noticing that for every $x \in N$, every $y \in N$ is {\bf conjugate}
as the $G$-action produces families of geodesics.

\paragraph{}
If $M$ admits an isometry group $G$, by Hodge, the harmonic forms representing cohomology
classes are $G$-invariant. They can live also on the fixed point set $N$. For 
the flag manifold $M=W^6$ admitting a $G=U(1)$ action, where 
$N=\mathbb{S}^2 + \mathbb{S}^2 + \mathbb{S}^0$, the two harmonic cohomology 
classes in the sector of differential $2$-forms live on two copies of $\mathbb{S}^2$. 
They are represented by {\bf algebraic cycles} which here are manifolds even. 

\paragraph{}
How do we get from $W^6$ to $W^{12}$? We can think of passing over $\mathbb{CP}^5$
by first making an $G=SU(2)$ extension, then do a $G=U(1)$-extension. But the later is not possible
because of Grove-Searle theorem \cite{GroveSearle} which
forces for co-dimension $2$ extensions $N \to M$ that $N$ is a sphere or projective space. 
What happens is that the $U(1)$-symmetry on $W^{12}$ directly gives the fixed point set 
$N=W^6$ or $E^6$. Also seeing $N=\mathbb{CP}^2 + \mathbb{CP}^2$ as a $G=U(1)$ fixed point set of $M=W^6$ is 
not possible by Frankel but despite that $N=\mathbb{S}^2+\mathbb{RP}^0 \to M=\mathbb{CP}^2 + \mathbb{CP}^2$ is.
But $N=\mathbb{S}^2 + \mathbb{S}^2 + \mathbb{S}^0$ can be a $SU(2)$ fixed point set of $M=W^6$.
The extension fills $M-N$ by 4-spheres $\mathbb{S}^4$. 

\paragraph{}
For $W^6,E^6$ we have two harmonic $2$-forms and two harmonic $4$-form,
for $W^{12}$ we have two harmonic $4$ and two harmonic $8$-forms 
and for $W^{24}$ we have two harmonic $8$ and two harmonic $16$-forms. 
Their {\bf cup products} could produce interesting numbers, when integrated over $M$.
This still needs to be computed. Compare with the harmonic $2$-forms of $\mathbb{T}^2$
with Betti vector $b(\mathbb{T}^2)=(1,2,1)$. 
The lengths of the two cycles depends on the metric but they always intersect in 2 points. 

\paragraph{}
The association with mass appears because
the gauge bosons, photon, gluon or graviton all have zero mass. 
The cohomology of projective
spaces is different. For $\mathbb{RP}^{2d}$ it is always the Betti vector $b=(1,0,0, \dots )$. 
For $\mathbb{CP}^d$ it is $(1,0,1,0\dots,0,1,0,1)$ for 
$\mathbb{HP}^d$ it is $(1,0,0,0,1,0,0, \dots,1,0,0,0,1)$ and
for $\mathbb{OP}^1$ it is $(1,0,0,0,0,0,0,0,1)$  and
for $\mathbb{OP}^2$ it is $(1,0,0,0,0,0,0,0,1,0,0,0,0,0,0,0,1)$. 
The concrete question therefore is whether $M$ a Betti number $2$ or more
can be linked to {\bf transport properties} and so with mass. 

\paragraph{}
This could happen by some kind of interaction
between the different cohomology classes of the same kind. The
picture is not unfamilar when looking at {\bf ground states} and especially {\bf harmonic forms} 
have always been important in physics. Here, these forms appear as $2$ and $4$ forms
$u_2,v_2,u_4,v_6$ in $6$ manifolds $W^6,E^6$ and can define, an intersection forms
$dV=u_2 \wedge v_6$. In the $12$-manifold $W^6$ we can look at pairs $dV=u_4 \wedge v_8$
and in $W^{24}$ at pairs $dV=u_8 \wedge v_{16}$. It would be good to be able to
compare the numbers $\int_M dV$ with the masses of the particles. 

\paragraph{}
For $M=W^6$ for example, one has two different homotopically
nontrivial 2-spheres embedded in $M$ and one can ask for the minimal 
volume or notions of distance which are topologically robust.
This motivates to compute more quantitative numbers coming out of even-dimensional 
positive curvature manifolds. The {\bf volumes of cup products of cohomology classes} could be 
computed using the heat flow $e^{-L t}$ using the Hodge Laplacian $L=(d+d^*)^2$ 
which keeps every space $\Lambda_k$ of $k$-forms invariant.  
It would not surprise that the {\bf spectra} of the $L_k$
have a relation with the mass ratios. One would be able to ``hear the mass".

\paragraph{}
Speaking about Laplacians, one always can look at the {\bf spectrum of the 
Laplacians} $L_k$ on $k$-forms. Of special interest are the {\bf ground states}, 
the smallest non-zero eigenvalues in each $k$-sector. In the case of trivial 
cohomology, one can look at the smallest eigenvalue. 
By McKean-Singer \cite{McKeanSinger}, there is a symmetry between non-zero 
eigenvalues of even differential forms and non-zero eigenvalues of odd-dimensional differential forms. 
This fact could help to compute the ground states. It would be of extreme 
importance to know the lowest eigenvalues. Speaking about spectral invariants, 
also the {\bf Zeta regularized determinants}
are of interest. See e.g \cite{Osg+88,LapidusFrankenhuijsen}.

\paragraph{}
An other source of rich interest are the {\bf automorphism groups} of the manifolds. 
That is especially interesting for $\mathbb{OP}^2$ and $\mathbb{W}^{24}$, where
the symmetry group is $F_4$ one of the exceptional Lie groups
$G_2,F_4,E_5,E_7,E_8$. This Lie group $F_4$ has dimension $26$.
So, when looking for any numbers which make up the mass, one should also look at 
properties of these Lie groups. Rather than doing numerology, it would be 
better to see whether there are physical reasons why these manifolds at all should
appear in transport. 

\paragraph{}
One has not observed a single example of an even-dimensional compact positive curvature manifold, 
where not some metric with a symmetry exists. A continuum symmetry forces geodesics to come in 
families away from the fixed point set $N$ 
(a picture familiar in string theory, where particles are loops, but where also the variational
problem changes from arc length to area \cite{Albeverio1997}). 
The presence of a symmetry in a positive curvature manifold $M$
always has a non-empty fixed point set $N$ by a theorem of Berger \cite{Berger1961}
which can be seen geometrically: a geodesic segment in $M$ away from a fixed point set 
$N$ longer than the diameter of $M$, because of the focusing nature of the geodesic flow 
acting on families of geodesics governed by the {\bf Jacobi fields}.

\paragraph{}
It is intriguing that with a $G$-symmetry, $G=U(1),SU(2)$ any pair of {\bf fat geodesics} $G \gamma_1,G\gamma_2$ intersecting 
outside $N$ must outside $N$ intersect in sets $Gx$ which are topologically robust meaning that if we 
change some initial point or velocity of a geodesic, this number does not change.
The symmetry renders the intersection points to be either circles $U(1)$ or $3$-spheres $SU(2)$. 
One can study the interaction between such {\bf fat geodesics} $G \gamma$. If two such sets
$G \gamma_1, G \gamma_2$ intersect, they intersect in a sphere $G x$
The number of such intersection points for geodesics starting and ending in $N$ 
is an integer {\bf intersection number} that is topologically stable. 

\paragraph{}
The structure of {\bf odd-dimensional positive curvature manifolds} with symmetry 
is a bit more complicated \cite{Wolf2011}. We first have the {\bf space forms}
$\mathbb{S}^{2d+1}_m = \mathbb{S}^{2d+1}/\mathbb{Z}_m$ which include projective spaces 
$\mathbb{RP}^{2d+1} = \mathbb{S}^{2d+1}/\mathbb{Z}_2$. The analog of $W^6$
are  the {\bf Aloff-Wallach spaces} $W^7_{p,q}$, which generalize
the {\bf Berger space} $B^7$. 
The analogue of $E^6$ are the Eschenburg spaces $E_{k,l}$. The analog of $W^{12}$ are the 
{\bf Bazaikin spaces} $B^{13}_q$ which generalize the {\bf Berger space} $B^{13}$. 
There appears no analog of $W^{24}$. Since the Euler characteristic is zero
for all odd-dimensional manifolds, one needs to organize therm using the 
fundamental group or look at Lefschetz numbers for some symmetry. 

\paragraph{}
Whether also affinities of Fermions with $(2d+1)$-manifolds of positive curvature can be drown 
is not at all clear. Unlike in even dimensions, where all known 
positive curvature manifolds have also symmetry, 
in odd dimensions, there are also positive curvature space forms without 
continuum symmetry. In odd dimensions, we also can have 
discrete symmetries $G$ which have maps $T:M \to M$ as generators.
In that case, one can look at the {\bf Lefschetz
number} $\chi(M,T)$ which is the super trace of the linear map induced on cohomology.
It is equal to $\chi(N)$ in general and $\chi(M)$ if $T$ is the identity.
For $M=\mathbb{S}^{2d+1}$ and a reflection $T:x \to -x$ for example, $\chi(M,T)=2$
and this is $\chi(N)=2$, the manifold $N=\mathbb{S}^{2d}$ fixed by $T$.

\paragraph{}
Is it possible to match up the Leptons ``electrons" and ``neutrini" or 
the six quarks {\bf up}, {\bf down}, {\bf strange}, {\bf charm}, 
{\bf top},{\bf bottom} with positive curvature $(2d+1)$-manifolds, where the later are
possibly equipped with some discrete symmetry $T: M \to M$? 
Inspired from the Harmonices Mundi analogy given by Kepler, one can note that 
{\bf space form} symmetries \cite{Wolf2011} can be {\bf ``tetrahedral"},
{\bf ``octahedral-cube"} like or {\bf ``icosahedral-dodecahedron"} like. 
A most naive analogy would try to match symmetry types with {\bf three quark and lepton generations}.

\paragraph{}
Using a discrete isometric symmetry $T:M \to M$ one can again place the 
objects in a {\bf dimension-Lefschetz number plane} as Euler characteristic $\chi(M)=\chi(M,Id)$ 
is always zero in odd-dimensions if Poincar\'e duality holds, but the {\bf Lefschetz number}
$\chi(M,T)$ is more informative. The non-orientable $\mathbb{RP}^{2d+1}$
with $\chi(\mathbb{RP}^{n})=1$ independent of dimension suggests a match-up
of real projective spaces with {\bf neutrini}. And the
{\bf dynamical spheres} $\mathbb{S}^{2d+1},T(x)=-x$ associate with {\bf electron-positron}
pairs. That would take care of the Leptons. The {\bf quarks} should then have to be matched up
with suitable pairs $(M,T)$ the {\bf Aloff-Wallach-Berger} and {\bf Bezaikin-Berger spaces} 
$M$ equipped with some discrete symmetry $T$. That no $25$-manifold analog of $W^{24}$ has
been detected in math is compatible with the absence of a Fermion like analog of the Higgs boson. 

\paragraph{}
The discrete symmetry also allows to associate some mass to electrons and neutrons
if one puts some symmetry on it. Euler characteristic can in nature be tied with
energy \cite{KnillEnergy2020}. Now, by taking space forms, the Euler characteristic
can become negative. One might associate odd cohomology $b_{2k+1}>1$ with negative
mass like anti-particles. For $2d$-manifolds, 
the statement $b_{2k+1}=0$ for all $k$ is stronger than the {\bf Hopf conjectures} \cite{Hopf1932}. 
Taking quotients changes Betti numbers according to spectral sequences:
for $SU(3)$ for example one has $b(SU(3))=(1,0,0,1,0,1,0,0,1)$ with $\chi(SU(2))=0$ 
while for $W^6=SU(3)/\mathbb{T}^2$ one has $b(W^6)=(1,0,2,0,2,0,1)$ with $\chi(W^6)=6$. 

\paragraph{}
Note that ``mass" is still not well understood. There are mechanisms for particles to acquire
mass via the Higgs field, but the {\bf neutrino mass} is not believed to come from a Higgs
mechanism. Indeed, positive neutrino mass is placed beyond the standard model. 
The cohomological interpretation with associating mass with higher Betti numbers
would allow to get mass by factoring out discrete symmetries or by
using fibrations. The change of flavor of neutrini would have to mean a {\bf symmetry change}.
Discrete symmetry changes can be explained already by {\bf harmonic 
analysis}. Waves are superposition of an even and an odd wave for example.
A function on a sphere is a superposition of {\bf spherical harmonics} which can have different
symmetries. For $2d$-manifolds of positive curvature, there is no interesting discrete symmetry 
beside $\mathbb{Z}_2$ which allows for {\bf charge} but not for {\bf flavors}: gauge bosons 
do not come in generations. This changes in odd dimensions: already on the 
circle, Fourier series shows different type of symmetries of the eigenfunctions $e^{i n x}$. 
The richer varieties of space forms in odd dimensions correlate with a richer variety of 
Fermions.

\paragraph{}
The analogy drawn urges to compute
the {\bf numerical values} coming from even and odd-dimensional positive curvature manifolds. 
We do not even known the Betti numbers $b_k(M)$ of all odd-dimensional $M$ and Lefschetz
numbers $\chi(M,T)$ for discrete transformations.
We have not seen any properties of the non-trivial cohomology classes, especially in the Wallach 
space $W^6,W^{12},W^{24}$ case. We miss {\bf quantitative numbers} like {\bf ground state energies} of the 
Hodge Laplcian of positive curvature manifolds for all $k$-forms as well as volumes of cup products
of cohomology classes. Computing these numbers and seeing them not match up with
masses is a way to find out to {\bf falsify the analogy} \cite{SeiffertRadnitzky}. 
We can compute a basis of the cohomology group for any simplicial complexes 
with a few lines of code, but for simplicial complexes which 
come from $6$, $12$ or $24$- dimensional manifolds, we can not even store the incidence 
matrices corresponding to the exterior derivatives. It needs new ideas to compute those.

\bibliographystyle{plain}

\end{document}